\documentclass{stablecore}

\usepackage{algorithm2e}
\usepackage{algorithmic}

\def\sharedaffiliation{%
\end{tabular}
\begin{tabular}{c}}
\begin{document}

%

\title{ Stable Cluster Core Detection in Correlated Hashtag Graph}

%
%
%
%
%

\numberofauthors{1}
    \author{
      \alignauthor Qinyun Zhu\\  
      \affaddr{Syracuse University}\\   
      \email{qzhu02@syr.edu}
      \sharedaffiliation
          }
\maketitle
\begin{abstract}
Hashtags in twitter are used to track events, topics and activities. Correlated hashtag graph represents contextual relationships among these hashtags. Maximum clusters in the correlated hashtag graph can be contextually meaningful hashtag groups. In order to track the changes of the clusters and understand these hashtag groups, the hashtags in a cluster are categorized into two types: stable core and temporary members which are subject to change. Some initial studies are done in this project and 3 algorithms are designed, implemented and experimented to test them. 
\end{abstract}

%
%
\keywords{Dynamic ,Hashtag Graph, Twitter, Cluster}

\section{Introduction}
Twitter is one of the most popular online social networking and microblogging service providers in the world. People share information using twitter in a format of status messages called tweets. Since most of these tweets are accessible by public, it has become a huge source of information about undergoing events, topics and people's activities. Hashtags are written as a combination of keywords, abbreviations or argots in order to track the tweets of corresponding events, topics or activities. Therefore, meaningful groups of hashtags can represent currently ongoing events, topics and reflect people's interests and opinions. Tracing the changes of these groups can further give researcher better ideas about mining and understanding people's interest and opinions. To better understand and track the changes, identifying the unchanged part of a cluster is very helpful.

Major work in this project is composed of following: 
\begin{itemize}
\item Based on the correlated hashtag graphs, active hashtag graphs in a daily basis were constructed to represents current active correlations of hashtags.  
 
\item The concepts of stable core and temporary members of the hashtag graphs are defined in order to better trace the changes of the correlated hashtag groups. The key task is to identify the stable cores which has persistent relationship among the hashtags while the temporary members are assumed to be more about transient interests and opinions which may change later.  

\item As an initial attempt of stable core detection, three algorithms are designed, implemented and tested. a) the Top-N algorithm identify the top N most closely related hashtags in a cluster. b) the Above-Average-Support method find relationships whose support score above the average support of the cluster. c) the Threshold based method view all relationship above a threshold as a part of the stable core.

\end{itemize}

\section{Background}\label{sec-background}

11 days of tweets are collected via twitter streaming API which randomly samples 1\% of incoming twitter status. The hashtags are extracted from the tweets. For the hashtags which appear in a same tweet, they are assumed to be contextual correlated. 

The dynamic correlated hashtag graph is defined as a series of static undirected graph $G^t=(V^t,E^t)$. $V^t$ is set of the vertexes in the graph, i.e. the set of hashtags found at time $t$. The edges of static graph at time $t$ are defined as $e^t=(v_{1}^t,v_{2}^t)$ where $e^t \in E^t$ and $v_{1}^t,v_{2}^t \in V^t$. The hashtags have edges between them when they are appear in the same tweets. The vertexes and edges in the graph are weighted by support scores, i.e. the number of users use these hashtags. $U(v^t)$ is the number of users use the hashtag $v^t$ and $U(e^t)$ is the number of users use the hashtags of $e^t$ together at time $t$. Another metric used to evaluate the correlation of the hashtags and weight the edges is Jaccard coefficient $J(v_{1},v_{2})=\mid v_{1} \cap v_{2}\mid/\mid v_{1} \cup v_{2}\mid$. Similar definitions of support score are also used for keyword correlated graphs in \cite{Agarwal:2012:RTD:2336664.2336671}. 

\begin{figure}
\centering
\epsfig{file=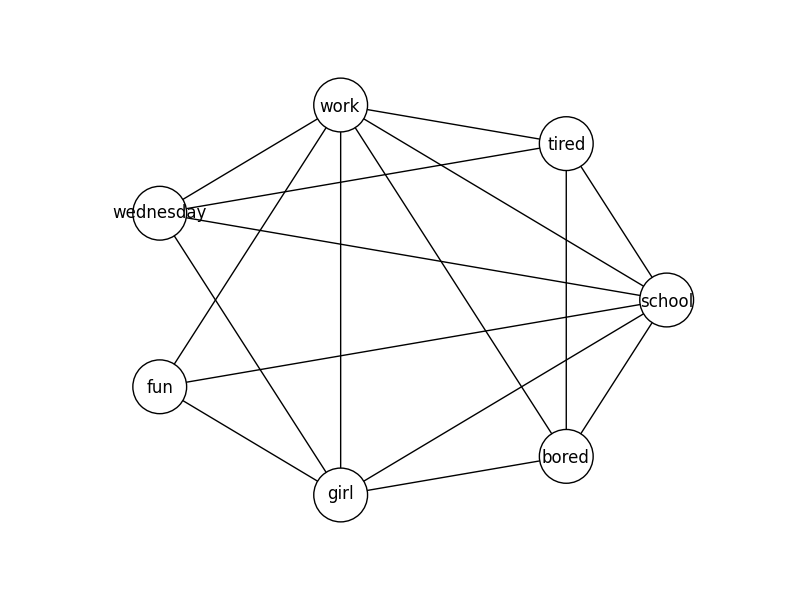,scale=0.3}
\caption{Clusters on Oct. 2nd}
\label{fig:SCHOOLWORK2}
\end{figure}

\begin{figure}
\centering
\epsfig{file=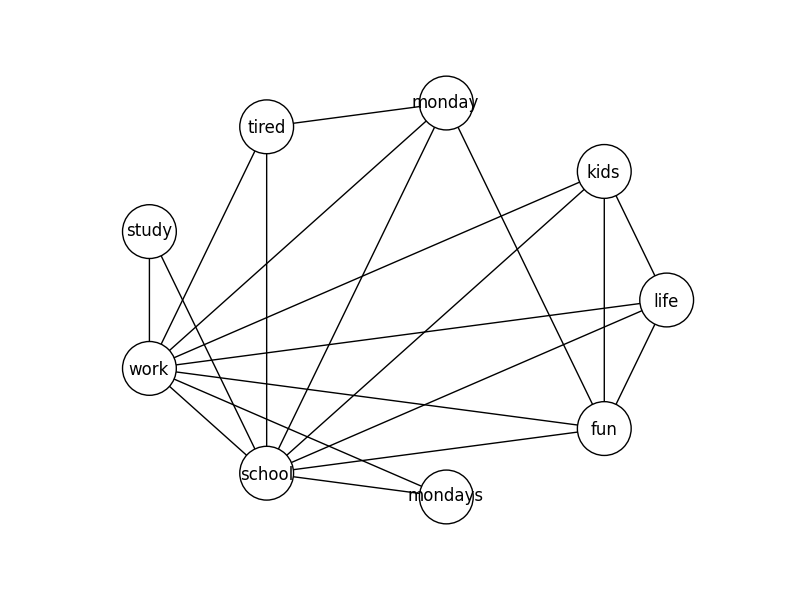,scale=0.3}
\caption{Clusters on Oct. 7th}
\label{fig:SCHOOLWORK7}
\end{figure}

The cluster in this project is defined as maximal clique \cite{Cazals2008564}. The clusters found by maximal clique detection algorithm can represent meaningful hashtag groups as examples in fig.\ref{fig:SCHOOLWORK2}, fig.\ref{fig:SCHOOLWORK7}. In the related works, changes of clusters are usually traced by community similarity/distance \cite{Takaffoli201149} \cite{DBLP:journals/corr/abs-1303-5009} or set relationship (super set, subset) \cite{anomalycomm2012}. The stable core studied in this project is proposed to be used as possible identity of the hashtag clusters. Fig.\ref{fig:SCHOOLWORK2} and fig.\ref{fig:SCHOOLWORK7} are examples of clusters based on 'work' and 'school' on different dates. Some members of the clusters are subject to change because they have temporary relationship with each other. For example, hashtags like 'Monday' and 'Wednesday' become active or inactive based on the weekday the hashtags are extracted. On the contrary, some hashtags like 'work' and 'school' have persistent correlations. Therefore, clusters' identity may be able to marked by the persistent part of the clusters plus some special ones from the temporary part. In this project, some exploration of finding the persistent part of the cluster, i.e. the stable core, has been done. 

\subsection{Active Correlated Hashtag Graph}
The active correlated hashtag graph (AHG) is a preprocessed graph of correlated hashtag graph. The hashtags and edges existing in the active graph are decided by the thresholds of hashtag support score $Thr_v$, edge support score $Thr_e$ and Jaccard coefficient of the edges $Thr_J$. So the AHG is defined as $G_{A}^{t} \subset G^t$ where for all $v^t$, $U(v^t) > Thr_v$ and for all $e^t$, $U(e^t) > Thr_e \wedge J(v_{1}^t,v_{2}^t)>Thr_J$ where $e^t=(v_{1}^t,v_{2}^t)$.  

\subsection{Stable Core}
The stable core is the persistent sub-cliques of a cluster. The clusters (i.e. maximal cliques) in AHG $G_{A}^t$ are $C(G_{A}^t)$. The stable cores $SC(C(G_{A}^t),N)$ are the sub-cliques of $C(G_{A}^t)$ that exist in the sub-cliques of $C(G_{A}^{t+N})$. For instance, there is a cluster, ["work","school","wednesday"], in $C(G_{A}^0)$ and there is a cluster in $C(G_{A}^{5})$, ["work","school","monday"]. One of stable cores is ["work","school"] in $SC(C(G_{A}^0),5)$.

\section{Method}\label{sec-method}

Intuitively, the more people support the usage of certain two hashtags together, the more the relationship is likely to exist for a longer time. Three approaches of finding stable cores based on the support score of edges in snapshot of AHG are explored. The input to these approaches is a cluster and output is one stable core of the input cluster.

\subsection{Top-N}
\begin{figure}
\textbf{TOP-N\textunderscore STABLE\textunderscore CORE:}

\begin{algorithmic}
\STATE {Input: Cluster $C$, Core size $N$}
\STATE {Output: $Core$ of $C$}
\STATE $Core \leftarrow \emptyset$
\IF {$len(C)>=N$}
\STATE {$Core \leftarrow \{$Hashtags in Top\_Scored\_Edge$(C)\}$}
\WHILE {$len(Core) < N$}
\STATE {$Core = Core \cup \{tag\}$ where $tag \in \{C \setminus Core\} \wedge tag \in $ Top\_Scored\_Edge$(Neighbors(Core) \cap \{C \setminus Core\})$}
\ENDWHILE
\ENDIF
\end{algorithmic}

\caption{Top-N stable core detection}
\label{fig:TOPN}
\end{figure}
The Top-N method first finds the top scored edge in the input cluster as the initial stable core. Then find the top edges left in the cluster connecting to hashtags in the stable core repeatedly until N hash tags found for the stable core. The method is described in fig.\ref{fig:TOPN}. The size of the stable cores are fixed in the method but the absolute support scores of these edges are not restricted. 

\subsection{Above Average Support}
The Above Average Support (AA) also starts from the top scored edge in the input cluster. Then the hashtags, which have edges with support score above the average of the cluster to all of the hashtags in the core, are added into the stable core. The method is described in fig. \ref{fig:AAS}. The average support $AvgSup(c)$ is the average support of clique $c$. The size of the stable cores are not fixed but the absolute support scores of these edges in the sable cores are still not restricted in this method.
\begin{figure}
\textbf{ABOVE\textunderscore AVERAGE\textunderscore CORE:}
\begin{algorithmic}
\STATE {Input: Cluster $C$}
\STATE {Output: $Core$ of $C$}
\STATE {$Core \leftarrow \{$Hashtags in Top\_Scored\_Edge$(C)\}$}

\FORALL {$tag \in C$}
\IF {$tag \notin Core \wedge (\forall c \in Core:Sup(tag,c)\geq AvgSup(C))$}
\STATE $Core \leftarrow Core \cup \{tag\}$
\ENDIF
\ENDFOR

\end{algorithmic}
\caption{AA stable core detection}
\label{fig:AAS}
\end{figure}

\subsection{Edge Threshold}
In this method as described in fig. \ref{fig:ET}, only the edge support threshold is considered. This approach is also started from the top supported edge and the process is very similar the AA method. The only difference is that instead of the average support of the cluster, a fixed edge support is used to decide which hashtag will be added into the core.

\begin{figure}
\textbf{EDGE\textunderscore THREASHOLD\textunderscore CORE:}
\begin{algorithmic}
\STATE {Input: Cluster $C$, Threshold $Thr$}
\STATE {Output: $Core$ of $C$}
\STATE {$Core \leftarrow \{$Hashtags in Top\_Scored\_Edge$(C)\}$}
\IF{$AvgSup(Core) \geq Thr$}

\FORALL {$tag \in C$}
\IF {$tag \notin Core \wedge (\forall c \in Core:Sup(tag,c)\geq Thr)$}
\STATE $Core \leftarrow Core \cup \{tag\}$
\ENDIF
\ENDFOR
\ELSE
\STATE{$Core \leftarrow \emptyset$}
\ENDIF

\end{algorithmic}
\caption{Edge Threshold}
\label{fig:ET}
\end{figure}

\section{Experiments}
The data was crawled from twitter between 10/2/2013 and 10/12/2013. The snapshots of the dynamic graph are taken each day. For each of the stable core detection methods, an experiment based on the first day of our data was done to find how well they can figure out the stable cores in the following days. The performance of the methods are evaluated by the real stable core ratio of $SC(G_{A}^0,N)$, which is the rate of the cores found on $G_{A}^0$ surviving on the $N$th day, i.e. $\mid G_{A}^0\mid /(\mid G_{A}^N\mid \cap \mid G_{A}^0\mid)$. Then the 3 methods were compared based on general real stable core ratio and amount of cores found.
\subsection{Performance of Top-N} 
\begin{figure}
\centering
\epsfig{file=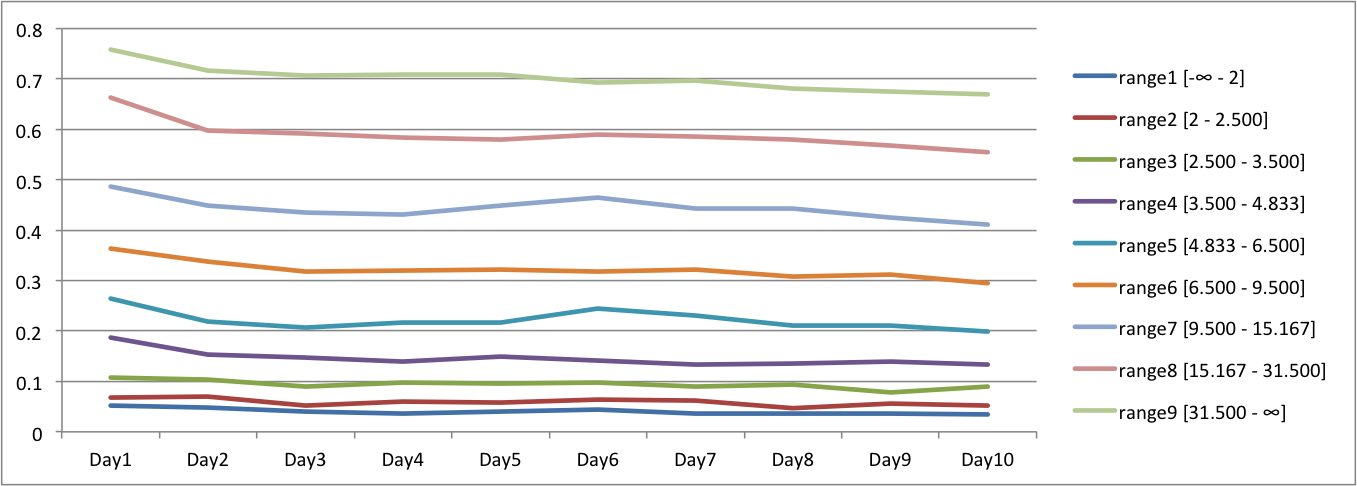,scale=0.37}
\caption{Real Core Ratios of Top-N}
\label{fig:RCTPN}
\end{figure}

The N in Top-N is configured to 3 in this experiment. The stable core ratios of $SC(G_{A}^0,n)$ where $n \in [1,10]$ are calculated. Each of lines shown in fig.\ref{fig:RCTPN} stands for the changes of real stable core ratio of cores having a specific range of average support score. The x-axis is for days and the y-axis is for the stable core ratio. The higher average support score the cores have, the higher ratio they will still active from 1st to 10th day after the beginning day. As the day progresses, the fraction of survived cores goes down when the average support score is relatively high. However, for the relatively low average support score, the ratio does not change much--they all are equally low. Therefore, the cores found using this method with high average support score are more likely to be real stable cores.

\subsection{Performance of AA}
\begin{figure}
\centering
\epsfig{file=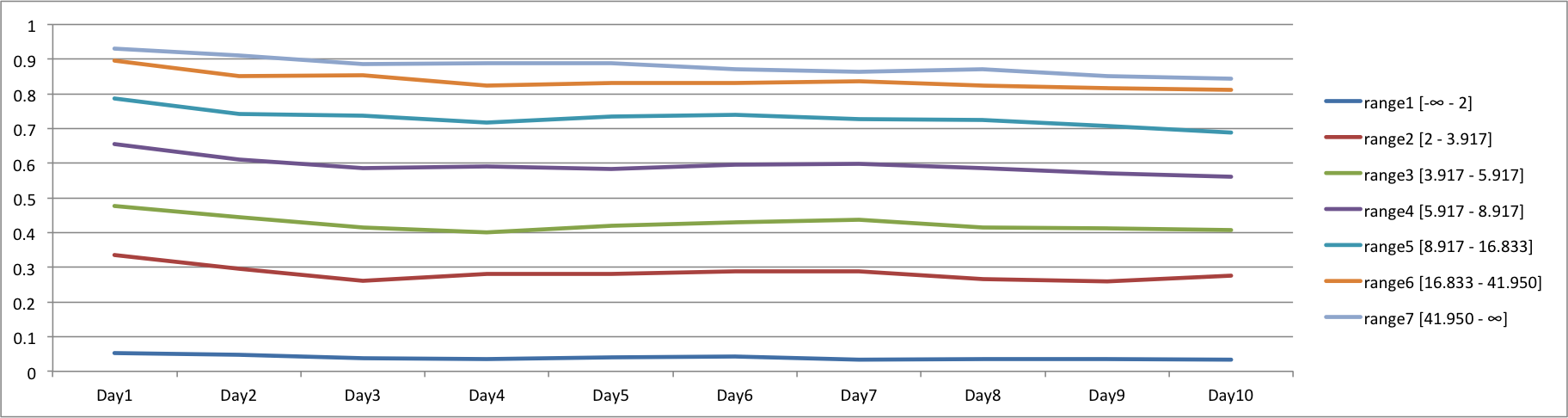,scale=0.3}
\caption{Real Core Ratios of AA}
\label{fig:RCAA}
\end{figure}

From fig.\ref{fig:RCAA}, we can see the pattern of performances of above average support method is very similar to the Top-N methods excepting that the highest stable core ratio in the top range of average support scored cores is higher.

\subsection{Performance of Edge Threshold}
\begin{figure}
\centering
\epsfig{file=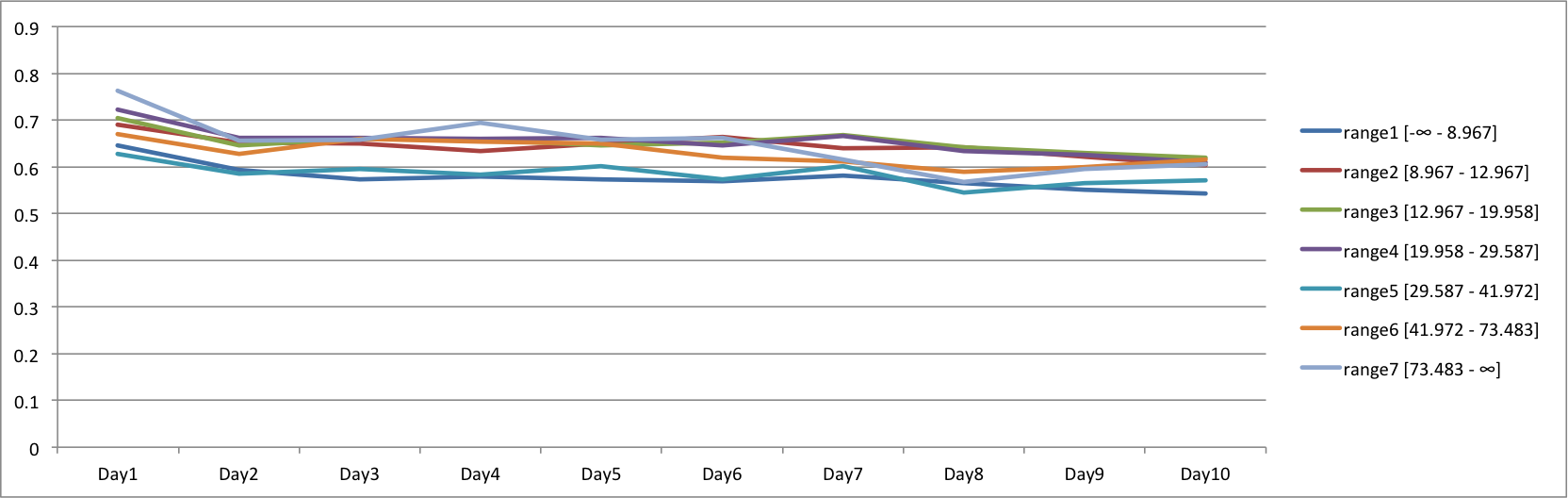,scale=0.3}
\caption{Real Core Ratios of Edge Threshold}
\label{fig:RCTHR}
\end{figure}

The performance of the edge support threshold method is shown in fig.\ref{fig:RCTHR}. From the observations of previous experiments, we found when the average support score of the cores is above 6, the real stable core ratio is above 0.6. So in this experiment, the edge support threshold is set to 6. The numbers fluctuate more than those compared to the previous two methods. It is probably because the edges in a core may have relatively low support score even if the average support score is high.  

\subsection{Comparison}
\begin{figure}
\centering
\epsfig{file=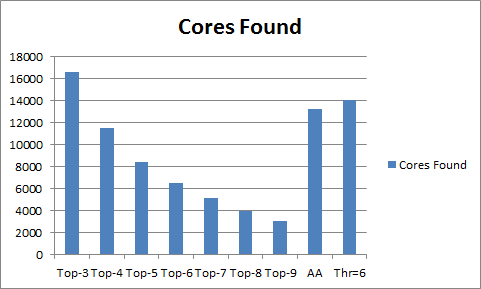,scale=0.65}
\caption{Cores Found}
\label{fig:CFC}
\end{figure}

\begin{figure}
\centering
\epsfig{file=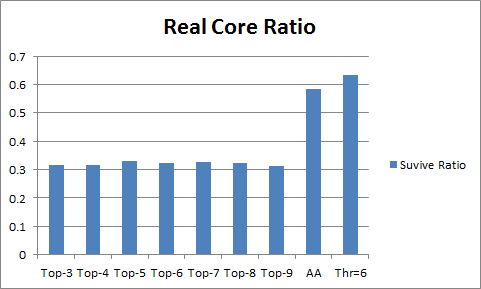,scale=0.65}
\caption{Real Core Ratios}
\label{fig:RCR}
\end{figure}

\begin{figure}
\centering
\epsfig{file=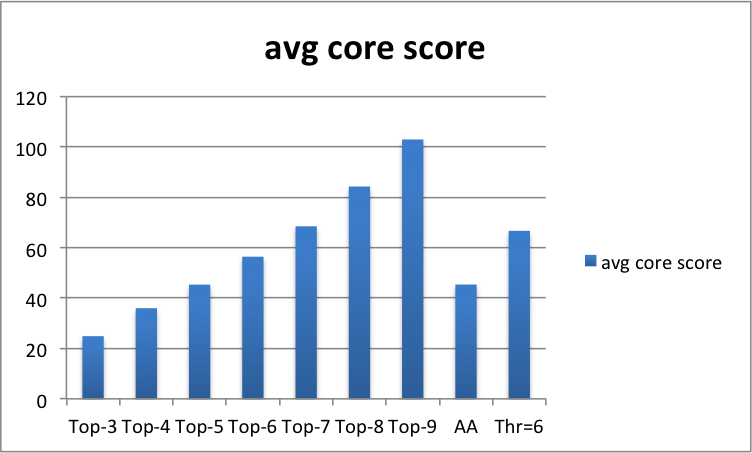,scale=0.65}
\caption{Average Core Score}
\label{fig:ACS}
\end{figure}

The performance is evaluated by real cores found and real core ratios. The high scores in these two metrics mean better performance. These two number relates to the recall and precision of the stable core detection method. The recall (ratio of total stable cores found by our algorithms) can not be easily calculated because we can not enumerate all the stable cores in such a large data set as twitter. But recall is directly proportional to the number of stable cores found by these algorithms. So we use the amount of stable cores to compare between different algorithms. In the comparison experiment, the methods are evaluated for detecting $SC(G_{A}^0,7)$. The fig.\ref{fig:CFC} shows the cores found by Top-N method with different configuration, AA methods and edge threshold methods with threshold set to 6. The amount of cores found by the Top-N method is the most among these evaluated methods but the fig.\ref{fig:RCR} shows that the real core ratios of all Top-N methods tested are low. The above average support method and edge support threshold method can find more cores than the Top-N methods excepting Top-3. The AA and edge threshold have much higher real core ratio than the Top-N ones. Thus, based only on these two metrics, the AA and edge threshold have better performance than the Top-N methods. Also, according to these figures, edge support threshold method with threshold=6 has slightly higher value in these two metrics comparing to the AA method. The fig.\ref{fig:ACS} shows that although Top-N with high N values has high average core score but these big ones still tend to split or disappear in the future. It is probably because some edges with relative low support are also included in the stable cores, which leads to unstable.

\section{Conclusion and Future Work}
From the above discussion, the Above Average support method and Edge Threshold method are more promising than the Top-N method. AA can find stable cores with very high precision in terms of real core ratio when average core score is high. Edge threshold method has a slightly general higher performance but may be more easily involve some edges with relatively lower than the average support scores in the cluster, which may be a reason why it has a slightly poor performance comparing to AA when the average score of the core is high. Therefore, combing these two methods may lead us to a better real core ratio but, to find stable cores as most as possible, more studies and thinking are needed. In order to improve both of the real core ratio and amount of cores detected, more factors like amount of clusters a core belonging to, historical appearance of a core, support changes of a edge should be taken into consideration. Also, more powerful prediction methods are needed to try. Also, metrics for quality of the cores found need to be better defined, how informative of a core is, what is the proper size of the cores need to be studied. Another issue is multiple stable cores may exist in a cluster. Current approach assumes that one stable core per cluster. Additionally, selection of temporary member for cluster identity is another part of work to complete the cluster identity and tracing problem. Not only be used to track changes as identity of the hash tag clusters, distinguishing the stable part and temporary members of the clusters can be a part of applications like opinion mining and hash tag recommendation system because both of the stable members and temporary members should be considered accordingly to reflect different aspects and temporary popular topics/events of relative hashtags.

%
\bibliographystyle{abbrv}
\bibliography{stablecore}  
%
%

\end{document}